\def\papertitle{InstructFX2FX: A Multi-Turn Text-to-Effect System for Sequential Audio Effect Refinement}
\def\paperauthorA{Song-Ze Yu}
\def\paperauthorB{Milan Liessens Dujardin}
\def\paperauthorC{Yuxuan Cai}
\def\paperauthorD{Wantong Zhang}
\def\paperauthorE{Brian Cruz}
\def\paperauthorF{Jeremy Wagner}
\def\paperauthorG{Carmine-Emanuele Cella}

\documentclass[twoside,a4paper]{article}
\usepackage{etoolbox}

\usepackage{dafx26demo}

\renewcommand{\permission}{%
 \if@blind
  \begin{figure}[b]
   \vspace{-.15cm}
   {\footnotesize
    \noindent\rule{.4\columnwidth}{0.4pt}\\[3pt]
    Copyright:~\copyright\,2026 the Anonymous Authors. This is an open-access article distributed under the terms of the \href{http://creativecommons.org/licenses/by/4.0/}{Creative Commons Attribution 4.0 International License}, which permits unrestricted use, distribution, adaptation, and reproduction in any medium, provided the original author and source are credited.}
  \end{figure}
 \else
  \begin{figure}[b]
   \vspace{-.15cm}
   {\footnotesize
    \noindent\rule{.4\columnwidth}{0.4pt}\\[3pt]
    \ifnum\value{numauth}=1
     Copyright:~\copyright\,2026 \paperauthorA. This is an open-access article distributed under the terms of the \href{http://creativecommons.org/licenses/by/4.0/}{Creative Commons Attribution 4.0 International License}, which permits unrestricted use, distribution, adaptation, and reproduction in any medium, provided the original author and source are credited.%
    \else
     Copyright:\,\copyright\,2026 \paperauthorA\ et al. This is an open-access article distributed under the terms of the \href{http://creativecommons.org/licenses/by/4.0/}{Creative Commons Attribution 4.0 International License}, which permits unrestricted use, distribution, adaptation, and reproduction in any medium, provided the original author and source are credited.%
    \fi}
  \end{figure}
 \fi
}

\usepackage{amsmath,amssymb,amsfonts,amsthm}
\usepackage[utf8]{inputenc}
\usepackage{ifpdf}
\usepackage[english]{babel}
\usepackage{caption}
\usepackage{subfig}
\usepackage{color}
\usepackage{booktabs}
\usepackage{tabularx}

\input glyphtounicode
\ninept

\newcounter{numauth}
\newcounter{listcnt}
\newcommand\authcnt[1]{\ifdefined#1 \stepcounter{numauth} \fi}

\newcommand\addauth[1]{
\ifdefined#1
\stepcounter{listcnt}
\ifnum \value{listcnt}<\value{numauth}
\appto\authorslist{, #1}
\else
\appto\authorslist{~and~#1}
\fi
\fi}
\authcnt{\paperauthorB}
\authcnt{\paperauthorC}
\authcnt{\paperauthorD}
\authcnt{\paperauthorE}
\authcnt{\paperauthorF}
\authcnt{\paperauthorG}
\authcnt{\paperauthorH}
\authcnt{\paperauthorI}
\authcnt{\paperauthorJ}
\def\authorslist{\paperauthorA}
\addauth{\paperauthorB}
\addauth{\paperauthorC}
\addauth{\paperauthorD}
\addauth{\paperauthorE}
\addauth{\paperauthorF}
\addauth{\paperauthorG}
\addauth{\paperauthorH}
\addauth{\paperauthorI}
\addauth{\paperauthorJ}

\usepackage{times}

\newif\ifpdf
\ifx\pdfoutput\relax
\else
   \ifcase\pdfoutput
      \pdffalse
   \else
      \pdftrue
   \fi
\fi

\ifpdf 
  \usepackage[pdftex,
    pdftitle={\papertitle},
    pdfauthor={\authorslist},
    pdfsubject={Proceedings of the 29th International Conference on Digital Audio Effects (DAFx26)},
    hidelinks,
    bookmarksnumbered,
    pdfstartview=XYZ
  ]{hyperref}
  \usepackage[pdftex]{graphicx}
\else 
  \usepackage[dvips]{epsfig,graphicx}
  \usepackage[dvips,
    pdftitle={\papertitle},
    pdfauthor={\authorslist},
    pdfsubject={Proceedings of the 29th International Conference on Digital Audio Effects (DAFx26)},
    hidelinks,
    bookmarksnumbered,
    pdfstartview=XYZ
  ]{hyperref}
\fi
\usepackage[hypcap=true]{caption}

\title{\papertitle}

\affiliation
{\paperauthorA*, \paperauthorB, \paperauthorC, \paperauthorD, \paperauthorE, \paperauthorF~and~\paperauthorG*}
{\href{https://cnmat.berkeley.edu}{Center for New Music and Audio Technologies (CNMAT)} \\ University of California, Berkeley\\
{\tt <vaclis, milan.ld, yuxuancai1106, wtzhang, brian.cruz, jeremy.wagner, carmine.cella>@berkeley.edu}\\[0.1em]
{\footnotesize *\,Corresponding authors}
}

\begin{document}
\ifpdf
  \DeclareGraphicsExtensions{.png,.jpg,.pdf}
\else
  \DeclareGraphicsExtensions{.eps}
\fi

\maketitle

\begin{abstract}
We present \emph{InstructFX2FX}, a system for sequential audio effect refinement through multi-turn natural-language instructions. Existing text-to-effect systems are largely single-shot, mapping one textual descriptor to one preset. Real audio engineering is instead sequential: engineers refine an existing effect chain through successive instructions. This poses a stateful problem that single-shot systems do not address: given the current effect parameters state and a new instruction, update the sound while preserving what earlier instructions already achieved.
InstructFX2FX addresses this with a hybrid architecture that divides labor between a language model and CLAP-guided optimization. The LLM serves as a high-level planner that selects effects and proposes the initial parameter state, motivated by recent evidence that LLMs can outperform CLAP-based optimization for single-turn text-to-effect mapping~\cite{Doh:2025:LLM2Fx}; CLAP-guided optimization then refines the existing parameter state, providing a more stable and robust refinement mechanism than LLM reprompting.
In the demo, attendees drive a dry recording through successive natural-language instructions: after each turn, they choose how strongly the effect is applied, then issue the next instruction based on what still differs from the sound they intend.
In a preliminary evaluation on SocialFX-derived descriptor pairs, CLAP-guided refinement achieves lower DSP-feature MMD than an LLM+LLM initialize-then-reprompt baseline on 9 of 10 pairs. Trajectory analysis further shows that, for differentiable effects, optimization tends to gradually move the audio toward the new target while retaining the effects of the previous instruction, highlighting the potential for gradual refinement.
Audio demo and source code are available at
\url{https://instructfx2fx.vaclis.net} and \url{https://github.com/vaclisinc/InstructFX2FX}.
\end{abstract}

\section{Introduction}
\label{sec:introduction}

Audio effect design is rarely a one-shot process. In practice, audio engineers and producers establish a rough effect chain and then iteratively refine it through feedback such as ``make it brighter,'' ``add some weight,'' or ``ease off the reverb.'' Each instruction is interpreted relative to the current sound and parameter state, not as an independent request from scratch.

We call this problem \emph{sequential FX refinement}: each new instruction updates the existing effect chain and parameter state, so that the rendered audio moves toward the user's evolving intent while preserving prior instructions. This differs from conventional text-to-effect generation, where each prompt is processed independently. We formalize it in Section~\ref{sec:problem}.

Recent works address only the single-turn case. Text2FX and FxSearcher both optimize effect parameters against a CLAP objective, differing mainly in the optimizer: Text2FX uses gradient descent through differentiable effects from a random initialization~\cite{Chu:2025:Text2FX}, while FxSearcher uses gradient-free Bayesian optimization for non-differentiable plugins~\cite{Ki:2025:FxSearcher}. LLM-based systems instead map text directly to parameters: LLM2Fx claims that an LLM, drawing on pretrained knowledge and in-context examples, can outperform CLAP-based optimization~\cite{Doh:2025:LLM2Fx}, and a subsequent tool-calling variant further selects and orders the FX chain through chain-of-thought planning~\cite{Doh:2026:LLM2FxTools}; both, however, remain single-shot. None of these directly support sequential refinement, and na\"ively reprompting an LLM at each turn is unreliable: the model regenerates parameters non-deterministically, may perturb settings the user did not intend to change across iterations, and cannot hear the sound it is editing.

We therefore propose a hybrid design in which the system works like a single audio engineer, pairing a brain with an ear. The language model acts as the ``brain'': it selects which effects to apply and generates reasonable initial parameters. CLAP-guided optimization then acts as the ``ear'': it listens to the rendered result and iteratively adjusts the parameters to pull the audio embedding toward the user instruction. The LLM thus provides the starting point, and CLAP the refinement direction.

We realize these ideas in \emph{InstructFX2FX}, a multi-turn text-to-effect system for sequential audio effect refinement. This paper makes three contributions. First, we formulate sequential FX refinement as a stateful, multi-turn parameter-update problem, distinct from single-shot text-to-effect generation. Second, we build a system that routes each instruction to reuse, initialize, or jointly refine an effect chain, pairing LLM-based planning with CLAP-guided gradient descent and Bayesian optimization. Third, we provide an interactive demonstration with a preliminary evaluation on the SocialFX dataset.
\begin{figure*}[t]
    \centering
    \includegraphics[width=0.9\linewidth]{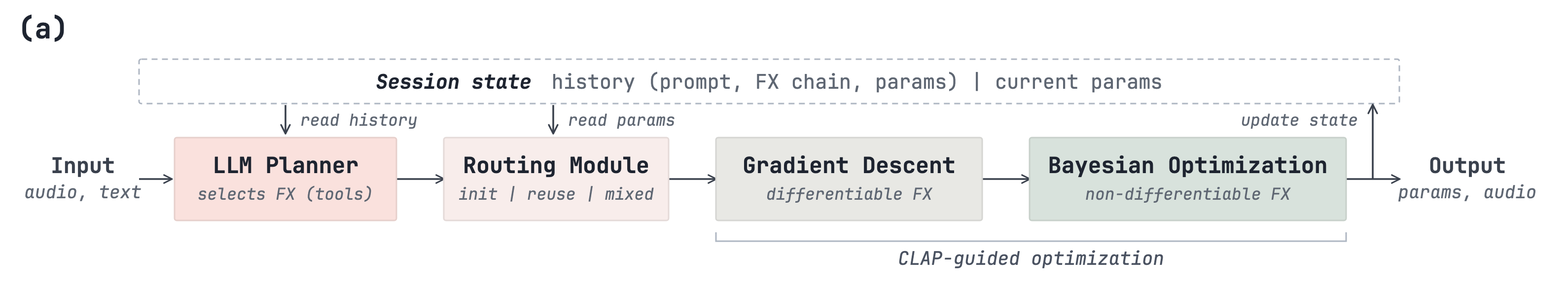}\\[2pt]
    \includegraphics[width=0.9\linewidth]{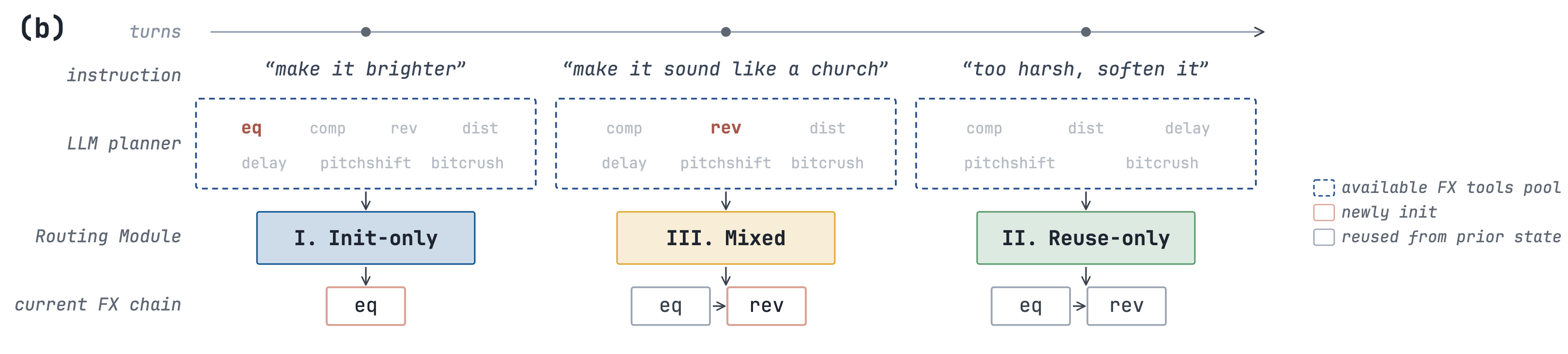}
    \caption{InstructFX2FX architecture. (a)~The harness: an LLM planner and routing module drive a CLAP-guided optimization backend over a persistent session state, which makes the system iterative. (b)~The three routing modes on an example three-turn session.}
    \label{fig:arch}
\end{figure*}

\section{Problem Formulation}
\label{sec:problem}

We formulate sequential FX refinement as a stateful parameter-update problem. Let $x$ denote the dry input audio, $C_t$ the current effect chain at turn $t$, $P_t$ the corresponding effect parameters, and $H_{t-1}=\{I_1,\ldots,I_{t-1}\}$ the prior instruction history. At each turn, given a user instruction $I_t$, the system produces an updated audio effects chain and parameter state through update function $f$:
\[
(C_t, P_t) = f(x, C_{t-1}, P_{t-1}, I_t, H_{t-1}).
\]
The instructions are applied sequentially and cumulatively across turns, denoted as $I_1 \rightarrow I_2 \rightarrow \ldots \rightarrow I_{t-1}$, where the right arrow reads as \textit{then}. 

By contrast, single-shot text-to-effect generation is defined as:
\[
P = g(x, I),
\]
where each instruction is processed separately, without tracking history. Sequential refinement, on the other hand, requires each instruction to move the current sound toward the new target while \textit{preserving} the effects of prior instructions. 
This requirement motivates the three design choices detailed in Section~\ref{sec:system}.

\begin{figure*}[t]
\centering
\begin{minipage}[c]{0.47\textwidth}
  \centering
  \subfloat[Exp 1: Sequential MMD across prompt pairs.]{%
    \includegraphics[width=\linewidth]{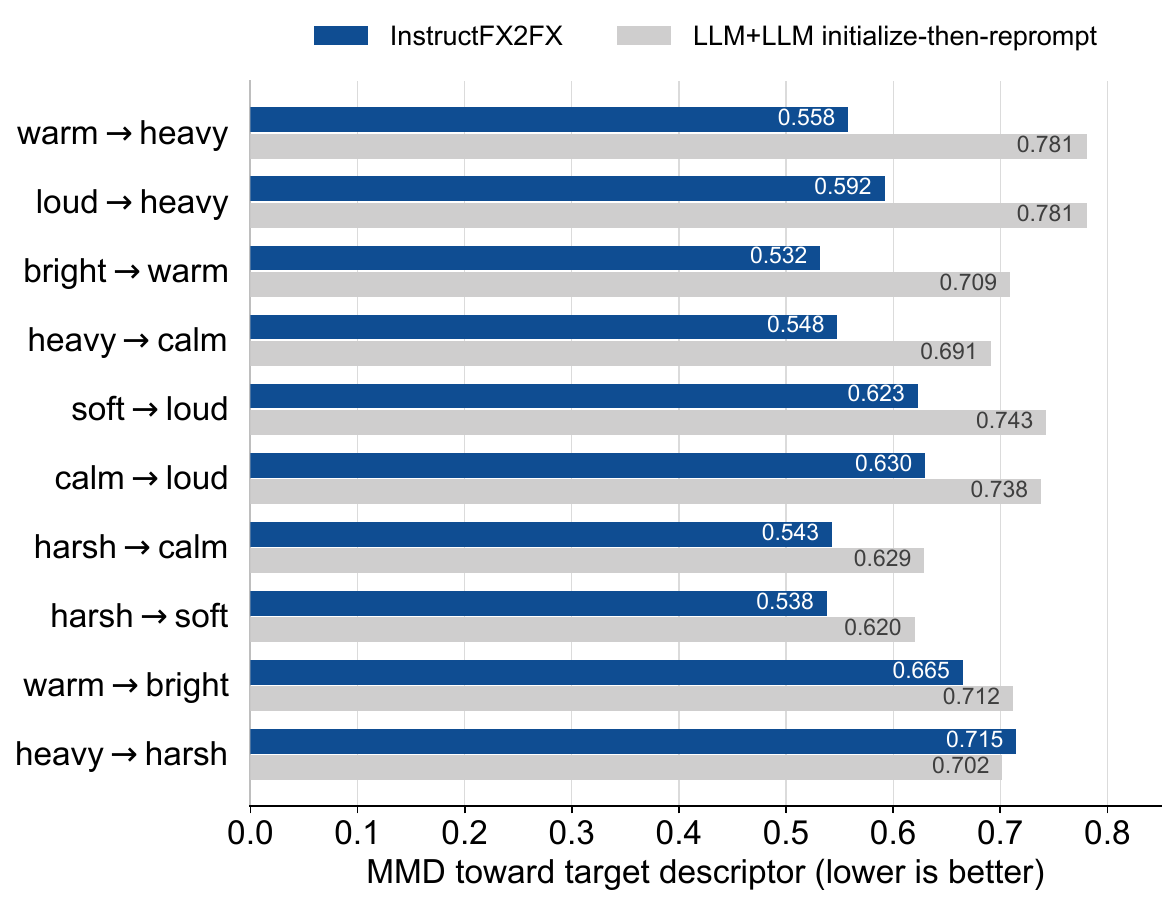}%
    \label{fig:exp1}}
\end{minipage}
\hfill
\begin{minipage}[c]{0.47\textwidth}
  \centering
  \subfloat[Exp 2: Case study of \emph{warmth} (init) — \emph{heaviness} (optimization target).]{%
    \includegraphics[width=\linewidth]{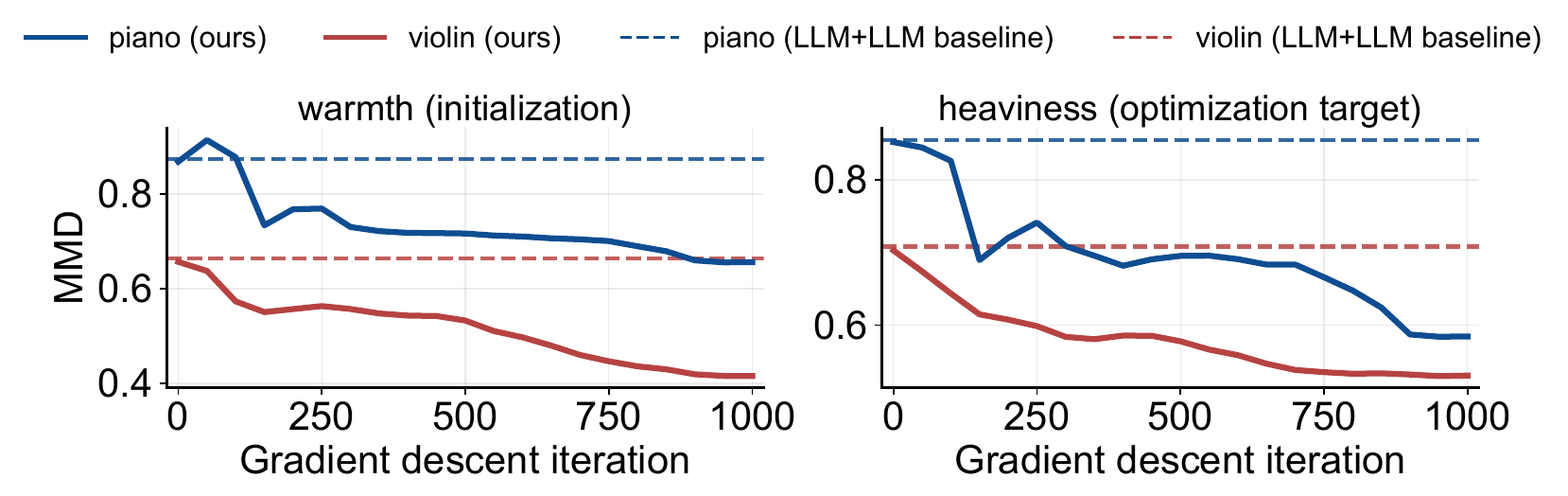}%
    \label{fig:exp2good}}\\[3pt]
  \subfloat[Exp 3: LLM initialization ablation.]{%
    \includegraphics[width=\linewidth]{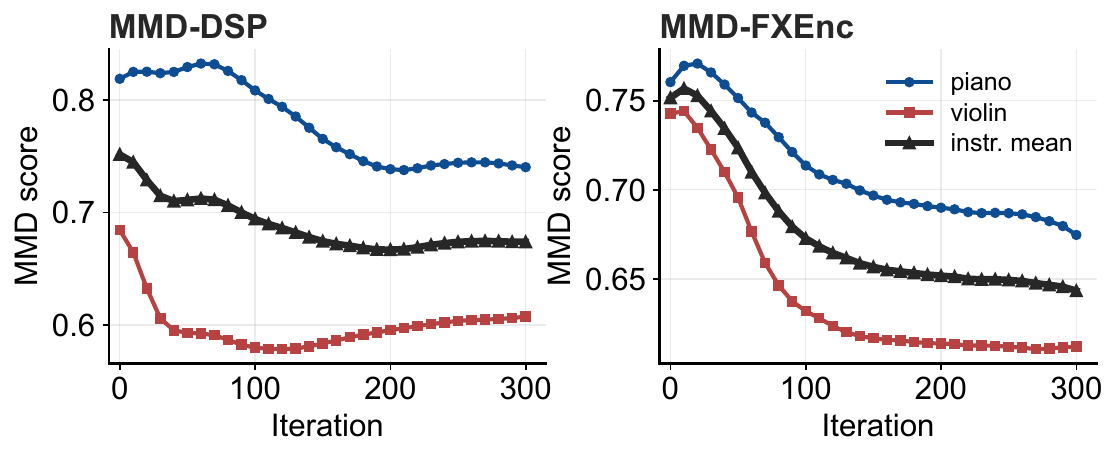}%
    \label{fig:exp5}}
\end{minipage}\\[3pt]
\caption{Evaluation results. (a) Exp 1: InstructFX2FX lowers MMD versus an LLM+LLM initialize-then-reprompt baseline on 9 of 10 pairs. (b) Exp 2: Refinement drives the target MMD (\emph{heaviness}, right) below the LLM+LLM initialize-then-reprompt baseline while the initialized MMD (\emph{warmth}, left) does not rise, showing that the audio moves toward the target without discarding earlier instruction. (c) Exp 3: departing from LLM initialization, CLAP-guided refinement achieves lower DSP-feature MMD (left) and Fx-Encoder++ MMD (right).}

\label{fig:results}
\end{figure*}

\section{System Overview}
\label{sec:system}

InstructFX2FX maintains a persistent session state so that the update function $f$ of Section~\ref{sec:problem} can build on earlier turns. As shown in Figure~\ref{fig:arch}(a), this state has two parts: a history of past turns, each recording the instruction, its effect chain, and the resulting parameters; and the current parameter state. The three components of the system each consult this state before acting: the LLM planner, the routing module, and the CLAP-guided optimization backend, which we detail below.

\subsection{LLM Planner}
\label{sec:planner}

The LLM planner serves as the system's decision layer, translating each natural-language instruction into a decision about which effects the updated chain should contain. Leveraging its pretrained audio-production knowledge, the LLM orchestrates the composition of the effect chain across turns, conditioned on the session state and history. Selection occurs over an explicit pool of supported effects: each effect is exposed as a callable tool, and effects already present in the session are withheld from the available tool set; if needed, they are refined in-place.

\subsection{Routing Module}
\label{sec:routing}

From the LLM planner's selection, the routing module compares the required effects against the current chain and dispatches the instruction into one of three modes:

\begin{itemize}
    \item \textbf{Initialize-only:} no existing effects are present, so the LLM initializes a new chain and parameters.
    \item \textbf{Reuse-only:} all relevant effects already exist, so the system refines the current parameters in place.
    \item \textbf{Mixed:} part of the existing chain is reused, while newly required effects are initialized and inserted.
\end{itemize}

Figure~\ref{fig:arch}(b) illustrates these three modes in an example three-turn session.

\subsection{Optimization Backends}
\label{sec:optimization}
Once the routing module has assembled the effect chain and parameter state, InstructFX2FX refines the parameter state through CLAP-guided optimization. We currently support two differentiable effects (EQ, reverb) and five non-differentiable Pedalboard effects (compressor, distortion, delay, bitcrush, pitch shift).

The two backends minimize CLAP-based objectives adapted from prior text-guided FX work: gradient descent for the differentiable effects and Bayesian optimization for the non-differentiable ones \cite{Chu:2025:Text2FX,Ki:2025:FxSearcher}. Let $\phi_\text{audio}(\cdot)$ and $\phi_\text{text}(\cdot)$ denote the CLAP audio and text embedding functions~\cite{Wu:2024:CLAP}, and let $\mathbf{x}$ denote the audio rendered by processing the dry input $x$ through the current effect chain.

The framework provides three interchangeable objectives, any of which can be selected per turn. The semantic similarity objective directly aligns rendered audio with the target instruction:
\begin{equation}
    \mathcal{L}_\text{sem} =
    1 -
    \cos\!\bigl(
    \phi_\text{audio}(\mathbf{x}),
    \phi_\text{text}(I)
    \bigr).
\end{equation}
For sequential refinement, the directional objective aligns the change in the audio embedding with the semantic transition from a source descriptor $S$ to a target descriptor $T$:
\begin{equation}
    \mathcal{L}_\text{dir} =
    1 -
    \cos\!\bigl(
    \phi_\text{audio}(\mathbf{x}(k)) -
    \phi_\text{audio}(\mathbf{x}_0),
    \phi_\text{text}(T) -
    \phi_\text{text}(S)
    \bigr),
\end{equation}
where $\mathbf{x}(k)$ is the rendered audio at optimization step $k$ and
$\mathbf{x}_0$ its value at initialization. 

Finally, a guided objective pushes the rendered audio toward a positive target $I^+$ and away from a negative anchor $I^-$:
\begin{equation}
    \mathcal{L}_\text{guided} =
    1 -
    \cos\!\bigl(
    \phi_\text{audio}(\mathbf{x}),
    \phi_\text{text}(I^+)
    \bigr)
    +
    \cos\!\bigl(
    \phi_\text{audio}(\mathbf{x}),
    \phi_\text{text}(I^-)
    \bigr).
\end{equation}

Because the desired strength of an effect is subjective, the system stores snapshots along optimization trajectories and exposes them in the demo interface as a slider, letting the user audition intermediate results and choose the effect strength they prefer.

\section{Evaluation}
\label{sec:evaluation}

We run a preliminary evaluation on EQ descriptors from the SocialFX dataset~\cite{Zheng:2016:SocialFX}, testing whether CLAP-guided refinement improves on both an LLM+LLM initialize-then-reprompt baseline and the LLM's own initialization.

\subsection{Dataset}
\label{subsec:dataset}

We evaluate on SocialFX~\cite{Zheng:2016:SocialFX}, a crowd-sourced
dataset that pairs natural-language descriptors with
audio-effect settings. We restrict attention to EQ descriptors and keep
those that are both frequent (at least 20 annotations) and reliable
(inter-annotator consistency at least 0.45), which yields seven words:
\emph{warm}, \emph{bright}, \emph{soft}, \emph{loud}, \emph{harsh},
\emph{calm}, and \emph{heavy}. From these we form 10 pairs of sequential instructions (e.g., \emph{warm}\,$\to$\,\emph{bright},
\emph{harsh}\,$\to$\,\emph{soft}), which we apply to dry self-recorded violin and open-source piano recordings.

\subsection{Metric: Maximum Mean Discrepancy}

We measure quality with the Maximum Mean Discrepancy (MMD) between the
system-output and ground-truth distributions, using a Gaussian kernel
(median-bandwidth heuristic) over 35-dimensional DSP feature vectors
(spectral, cepstral, and temporal descriptors); lower is better. Because
MMD is computed on DSP features, it is
independent of the optimization objective defined within the CLAP space, so a reduction in the MMD score compared to a baseline effectively shows that our system improves the manipulation of audio effects at the signal level. For Experiment~3, we
additionally report MMD in the Fx-Encoder++~\cite{yehFxEncoderExtractingInstrumentWise2025}
embedding space, a pretrained audio-effect encoder, used in prior work on single-shot parameter setting \cite{Doh:2025:LLM2Fx}. 

\subsection{Experiment 1: Sequential MMD}

For each of the 10 pairs of sequential instructions $A \to B$ described in Section~\ref{subsec:dataset}, we construct the ground-truth references by first applying the SocialFX EQ parameters of descriptor A to the dry audio, then stacking those of descriptor B on top. Experiment~1 compares the DSP-feature MMD of InstructFX2FX and an LLM+LLM initialize-then-reprompt baseline against this ground truth. Both methods begin with LLM initialization for descriptor $A$ . InstructFX2FX then refines the parameters according to the applicable routing scheme, whereas the baseline applies an LLM reprompt for descriptor $B$. We find that InstructFX2FX achieves lower DSP-feature MMD than the baseline on 9 of 10 descriptor pairs (Figure\ref{fig:exp1}). The largest improvement is observed for the strongest timbral transition, warm \,$\to$\,heavy (roughly 0.22). 

\subsection{Experiment 2: MMD Convergence Trajectory}

For Experiment~2, we build two independent ground truth references, complementing the sequential ground truth used in Experiment~1. Specifically, we apply the SocialFX EQ parameters of descriptors $A$ and $B$ \emph{separately} to the dry audio, and record the DSP-feature MMD of the refined audio to each single-word reference throughout CLAP optimization. A successful refinement should keep the MMD for the previous descriptor $A$ from increasing while reducing the MMD to the target descriptor $B$. Inspired by prior work \cite{Chu:2025:Text2FX}, we adopt the directional objective, where the source descriptor $S$ is chosen as the semantic opposite of the target descriptor $T$. This objective emphasizes directional alignment rather than direct similarity, encouraging refinement while retaining previous instructions. We observe this behavior for 9 of the 10 descriptor pairs across both piano and violin recordings. Figure~\ref{fig:exp2good} shows a representative example. Over CLAP gradient descent iterations, the DSP-feature MMD to the optimization target reference \emph{heavy} decreases, while the DSP-feature MMD to the initialization reference \emph{warm} does not rise, indicating that the optimization moves the audio toward the optimization target without discarding the previous instruction. Plots for all 10 descriptor pairs are available on the demo website. 

\subsection{Experiment 3: LLM Initialization Ablation}

This ablation isolates the contribution of CLAP-guided refinement from that of the LLM initialization. Unlike previous experiments, we evaluate on single-descriptor words: seven descriptors from Section~\ref{subsec:dataset} plus \emph{happy}. We obtain the ground truth references by applying the SocialFX EQ parameters of each descriptor to the dry audio. For each word and instrument, we sample the LLM at temperature~$1.0$ to obtain a set of stochastic initializations per pair. We then refine the parameter set by navigating the CLAP space toward the same target descriptor as the one used for initialization, making iteration~0 of every trajectory in Figure~\ref{fig:exp5} the LLM
baseline itself. Averaged across both instruments, refinement reduces DSP-feature MMD from 0.752 to 0.674 (about 10\%) and Fx-Encoder++ MMD from 0.752 to 0.643 (about 14\%), showing the value of CLAP optimization after LLM initialization for 7 of 8 words, the sole exception being \emph{warm}. 

Experiment 2 and 3 show that CLAP optimization offers a capability that LLMs alone cannot: gradual refinement. This strongly motivates the intermediate-checkpoint slider, allowing users to listen to gradual changes in the sound in order to ultimately settle on the parameter set that creates their desired effect.

\section{Limitations and Future Work}
\label{sec:limitations}

Several optimization and evaluation challenges remain. First, our quantitative evaluation centers primarily on EQ-related descriptors, because SocialFX provides cleaner descriptor-to-parameter associations for EQ than for more complex effects.

Second, CLAP embedding space does not capture all perceptual properties relevant to audio production~\cite{Deng:2025:Timbre}. Although CLAP-guided refinement often improves early behavior, trajectories may later drift or overshoot, and gains in CLAP space do not always correspond to gains in DSP-feature metrics or perceptual quality.

Third, non-differentiable effects remain difficult to refine reliably. Bayesian optimization extends support to Pedalboard effects such as distortion, delay, and bitcrushing, but its trajectories are often noisy and semantically inconsistent across iterations.

Finally, optimization currently requires seconds per interaction turn, limiting applicability within production DAW workflows.

Future directions include richer multi-effect chains, human listening studies, domain-adapted audio-effect embedding models, improving stability for non-differentiable refinement, and real-time VST/AU integration. Among these, the most promising in our view is to move beyond inference-time optimization toward generative parameter models, for example flow-matching models trained directly on iterative FX-editing trajectories that learn sequential refinement end-to-end.

\section{Conclusion}
\label{sec:conclusion}

We presented InstructFX2FX, a multi-turn text-to-effect system for sequential text-guided audio effect refinement. Unlike single-shot text-to-effect systems, InstructFX2FX treats audio effect editing as a stateful process in which each new instruction updates an existing chain and parameter state.

The core idea is leveraging complementary sources of machine intelligence: the LLM provides high-level effect planning and initialization, while CLAP-guided optimization provides perceptual refinement. Preliminary experiments on SocialFX-derived descriptors suggest that CLAP-guided refinement can improve over LLM reprompting in many sequential settings, while also exposing limitations such as CLAP/DSP-feature mismatch, optimization drift, and instability for non-differentiable effects. 
More broadly, this demo is a proof-of-concept 
for 
interactive audio-effect systems that better reflect practical creative workflows. We see it as a first step, and hope it motivates the community to take up \textit{sequential FX refinement} as a problem in its own right.

\bibliographystyle{IEEEtranDAFx}
\bibliography{instructfx2fx_demo}

\end{document}